\documentclass[
reprint,
nofootinbib,
amsmath,amssymb,
aps,
prl,
twocolumn,
floatfix,
]{revtex4-2}

\usepackage{amssymb}   
\usepackage{amsfonts}
\usepackage{amsmath}
\usepackage{mathtools}
\usepackage{MnSymbol}
\usepackage{dsfont}
\usepackage{dcolumn}   
\usepackage{bbm}
\usepackage{bm}        
\usepackage[mathscr]{eucal}
\usepackage{algorithmic}
\usepackage{algorithm}
\usepackage{placeins}
\usepackage{graphicx}
\usepackage[usenames,dvipsnames]{xcolor}

\usepackage{verbatim}

\usepackage{parskip} 

\usepackage{tikz}
\usetikzlibrary{quantikz}

\usepackage{float}

\usepackage{hyperref}
\definecolor{bblue}{RGB}{0, 126, 204}
\definecolor{dgreen}{RGB}{0,102,51}
\definecolor{blueus}{RGB}{10, 49, 97}
\definecolor{teal}{RGB}{0,128,128}
\definecolor{orangered}{RGB}{255,69,0}
\hypersetup{
	unicode=false,          
	pdftoolbar=true,        
	pdfmenubar=true,        
	pdffitwindow=false,     
	pdfstartview={FitH},    
	pdftitle={MPDO-ZNE},    
	pdfauthor={},     
	pdfsubject={},   
	pdfcreator={},   
	pdfproducer={}, 
	pdfkeywords={} {} {}, 
	pdfnewwindow=true,      
	colorlinks=true,       
	linkcolor=black, 
	citecolor=black,        
	filecolor=black,      
	urlcolor=black           
}

\DeclareMathOperator{\Tr}{Tr}

\newcommand{\vket}[1]{|{#1}\rrangle}

\newcommand{\eq}[1]{\begin{align}#1\end{align}}

\newcommand{\note}[1]{{\color{black} #1}}

\begin{document}

\title{Quantum simulation of dynamical phase transitions in noisy quantum devices}
\author{Younes Javanmard}
\email[]{younes.javanmard@itp.uni-hannover.de}
\author{ Ugne Liaubaite}
\email[]{ugne.liaubaite@itp.uni-hannover.de}
\author{Tobias J. Osborne}
\email[]{tobias.osborne@itp.uni-hannover.de}
\author{Luis Santos}
\email[]{santos@itp.uni-hannover.de}

\affiliation{Institut f\"ur Theoretische Physik, Leibniz Universit\"at Hannover, Appelstra{\ss}e 2, 30167 Hannover, Germany}

\date{\today}

\begin{abstract}
Zero-noise extrapolation provides an especially useful error mitigation method for noisy intermediate-scale quantum devices. 
Our analysis, based on matrix product density operators, of the transverse-field Ising model with depolarizing noise, reveals both advantages and inherent problems associated with zero-noise extrapolation when simulating non-equilibrium many-body dynamics. On the one hand, interestingly, noise alters systematically the behavior of the Loschmidt echo at the dynamical phase transition times, splitting the DQPT feature into a robust double-peak structure of the return rate, 
and hence inducing an error that, inherently, cannot be mitigated.  On the other, zero-noise extrapolation may be employed to recover quantum revivals of the Loschmidt echo, 
which would be completely missed in the absence of mitigation, and to retrieve faithfully noise-free inter-site correlations. 
Our results, which are in good agreement with those obtained using 
quantum simulators, reveal the potential of matrix product density operators for the investigation of the performance of quantum devices with a large number of qubits and deep noisy quantum circuits. 
\end{abstract}
\maketitle

Quantum computers may outperform their classical counterparts by exploiting the dimensionality of Hilbert space, which increases exponentially with the system size~\cite{nielsenQuantumComputationQuantum2010,feynmanSimulatingPhysicsComputers1982, lidar_brun_2013, lloydUniversalQuantumSimulators1996a}. 
They enable a large number of possible applications, including quantum simulation, optimization, machine learning, and more~\cite{biamonteQuantumMachineLearning2017, kandalaHardwareefficientVariationalQuantum2017, nielsenQuantumComputationQuantum2010}.
Despite major technical challenges, recent years have witnessed a rapid experimental progress on quantum computing using various platforms, including superconducting 
qubits~\cite{wendinQuantumInformationProcessing2017b}, trapped ion technologies~\cite{bruzewiczTrappedionQuantumComputing2019, ciracQuantumComputationsCold1995, blattQuantumSimulationsTrapped2012}, neutral atoms~\cite{jakschColdBosonicAtoms1998, blochQuantumSimulationsUltracold2012}, semiconductor spin qubits~\cite{burkardSemiconductorSpinQubits2021}, and photonics~\cite{slussarenkoPhotonicQuantumInformation2019, weedbrookGaussianQuantumInformation2012}.  

Currently available quantum information processing devices are severely limited by noise and decoherence. 
For this reason, they are referred to as noisy intermediate-scale quantum (NISQ) devices~\cite{preskillQuantumComputingNISQ2018}. 
These devices are limited to shallow quantum circuits, but, even then, noise can lead to faulty estimates. \note{Although quantum error correction can, in principle, reduce or eliminate the effect of decoherence it is not trivial to implement such schemes on present NISQ devices due to the limited number of qubits available, as well as the complex entanglement resource required.}

In response to the challenge of quantum noise, quantum error mitigation techniques have been developed for NISQ devices~\cite{temmeErrorMitigationShortDepth2017, endoPracticalQuantumError2018, liEfficientVariationalQuantum2017, I-Chi-Chen_Thomas-Iadecola-errormitigation2022} to ameliorate the deleterious effects of decoherence in a near-term compatible way. The two major general-purpose error mitigation methods are probabilistic error cancellation~\cite{bergProbabilisticErrorCancellation2022} and zero-noise extrapolation (ZNE)~\cite{liEfficientVariationalQuantum2017, mariExtendingQuantumProbabilistic2021, loweUnifiedApproachDatadriven2021}. 
Although there have been several studies on the general performance of mitigation techniques~\cite{takagiFundamentalLimitsQuantum2022, tsubouchiUniversalCostBound2022, 
takagiUniversalSampleLower2022, ladecolaPRR2022}, the behavior of ZNE for large-scale quantum simulations is not well explored. 

In this paper, we investigate, by means of 
matrix product density operators (MPDO)~\cite{verstraeteMatrixProductDensity2004a, zwolakMixedStateDynamicsOneDimensional2004a, chengSimulatingNoisyQuantum2021a, wernerPositiveTensorNetwork2016, SCHOLLWOCK201196}, the use of ZNE for the study of out-of-equilibrium quantum many-body systems.  
In particular, we evaluate the simulation of the dynamics of the transverse-field Ising model
by means of a NISQ device with depolarizing noise, characterized by a \note{local single-qubit depolarizing} noise channel \note{$\mathcal{E}(\rho) = (1-p)\rho + p\frac{I}{2}$} \cite{nielsenQuantumComputationQuantum2010, lidar_brun_2013}, with $\rho$ the density matrix, 
$p$ the probabilistic error rate that depends on the device
and the circuit, and $N$ the number of qubits. Although there are other possible noise channels, the depolarizing noise model often appropriately describes the average noise for large circuits involving many qubits and gates~\cite{urbanekMitigatingDepolarizingNoise2021a}. Our results indicate that, below a given noise and evolution-time threshold, ZNE may faithfully recover quantum revivals of the Loschmidt echo, which would be otherwise basically lost, as well as the inter-site correlation dynamics. However, we show that the noise-free dynamics during the time windows associated to dynamical quantum phase transitions (DQPTs) is inherently irretrievable using ZNE, due to a systematic noise-induced splitting of the DQPT feature into a double-peak structure of the return rate. Our results, which are in  good agreement with direct quantum 
simulations, show the potential of MPDO calculations for the study of NISQ devices with many qubits and deep circuits.


\paragraph{Model.--} 

We consider in the following the transverse-field Ising model ~(TFIM):
\eq{
\mathcal{H} = -\frac{1}{2}\left[ \sum^{N-1}_{i=1} \left (J_z  \sigma_i^z \sigma_{i+1}^z + J_x \sigma_i^x \sigma_{i+1}^x \right ) + \sum^N_{i=1}h_x \sigma_i^x \right], 
\label{eq: TFI}
}
where $\sigma_i^{x,y,z}$ are the spin-$\frac{1}{2}$ Pauli matrices at the different lattice sites $i=1, \dots, N$, $h_x$ is the transverse field, and 
$J_z$~($J_x$) denotes the $z$~($x$) spin-spin coupling. For convenience, we assume open-boundary conditions. 
For $J_x=0$, the model reduces to the integrable transverse-field Ising chain, which is exactly solvable by a Jordan-Wigner mapping into non-interacting fermions~\cite{pfeutyOnedimensionalIsingModel1970, Osborne-quantum-phase-transition2002}. 
To make the model generic, we add a weak $J_x$ term, which becomes equivalent to a two-particle interaction in the fermionic 
language, rendering the model non-integrable.  In this work, we consider $J_z=J$, 
$J_x=0.1 J$, and $h_x=0.1 J$. 

We are particularly interested in the non-equilibrium dynamics of Model~\eqref{eq: TFI} after a quench. 
Starting with a state $|\psi_0\rangle$, the Loschmidt echo is defined as the probability 
$\Lambda(t)=|\langle\psi_0 | e^{-i{\cal H}t} |\psi_0\rangle|^2$ to find the time-evolved system in the initial state. The return-rate function, 
$\lambda(t) = -\lim\limits_{N \to\infty}\frac{1}{N} \log \Lambda(t)$, is the real-time analogous of the free-energy per site. Non-analiticities of $\lambda(t)$ at critical times indicate the presence of DQPTs.
\note{DQPTs are a type of phase transition that occur in quantum systems as they evolve over time. Unlike traditional phase transitions that result from changes in external parameters, DQPTs are triggered by changes in the time-dependent Hamiltonian that governs the dynamics of the system. During a DQPT, there is a sudden change in the behavior of the system, which can be characterized by a variety of quantities, such as the entanglement entropy or the Loschmidt echo. These transitions can occur in a wide range of quantum systems and have been the subject of extensive research in recent years~\cite{heylDynamicalQuantumPhase2013, heylDynamicalQuantumPhase2014, heylDynamicalQuantumPhase2018, heylScalingUniversalityDynamical2015, karraschDynamicalPhaseTransitions2013a, dborinSimulatingGroundstateDynamical2022, HalimehDQPT-longrange2017, homrighausenAnomalousDynamicalPhase2017, langConcurrenceDynamicalPhase2018, vandammeAnatomyDynamicalQuantum2022, vandammeDynamicalQuantumPhase2022a, PhysRevD.106.094502}, which 
have been recently realized experimentally~\cite{jurcevicDirectObservationDynamical2017, wuDynamicalPhaseTransition2022}.
}


\paragraph{Master equation.--} 
In the presence of dissipation, the evolution of the density matrix $\rho$ of the system is given by the master equation~\cite{lidar_brun_2013, nielsenQuantumComputationQuantum2010}, 
\eq{ 
\partial_t \rho = & \mathcal{L}(t)[ \rho ] = -i[\mathcal{H}, \rho(t)]  \nonumber \\ & + \sum_{\substack{i\\\mu=x,y,z}}{\gamma_i \left( \sigma^\mu_i(t) \rho \sigma^{\mu\dagger}_i(t) -\frac{1}{2} \{\sigma^{\mu\dagger}_i(t) \sigma^\mu_i(t), \rho \} \right) }
\label{eq: Lindblad1}
}
where the first line provides the unitary part of the time evolution, whereas the second accounts for dissipation. 
The Lindblad~(jump) operators for depolarizing noise are given by the Pauli operators $\sigma_i^{x,y,z}$, and encode the system-reservoir couplings, $\gamma_i$ characterizes the dissipation strength at site $i$, and $[]$ and $\{\}$ denote commutator and anti-commutator, respectively.
For a time step $\delta t$, the probabilistic error is related to the noise strength through $p_{\delta t}= 1 - e^{-4\gamma \delta t}$~(see the Supplemental Material).
\note{For a Trotterized circuit with $M=t/\delta t$ steps on $N$ qubits, this corresponds to an effective accumulated circuit error probability $p_{\mathrm{circ}}(t)=1-(1-p_{\delta t})^{NM}\simeq NM\,p_{\delta t}\simeq 4N\gamma t$ for $p_{\delta t}\ll 1$.}
Assuming a constant $\gamma$ for each jump operator, we may re-write Eq.~\eqref{eq: Lindblad1} in the form:
\eq{
\partial_t \rho_\gamma(t) = {\cal L}[\rho_\gamma]=-i[\mathcal{H}(t), \rho_\gamma(t) ] + \gamma \mathcal{D}(\rho_\gamma(t))
\label{eq: Lindblad2}
}


\paragraph{Zero-noise extrapolation.--} 

ZNE is a relatively simple error-mitigation technique, easy to implement in post-processing, and which requires neither knowledge of the noise, nor a gate-level control of the device. The method assumes the ability to scale the noise parameter to particular values $\alpha \gamma$ with $\alpha>1$.
Assuming a constant noise rate, this may be achieved by a proper stretching of the time evolution by the factor $\alpha$, 
while reducing ${\cal H} \to {\cal H}/\alpha$, resulting in the same overall Hamiltonian evolution, whereas 
the noise increases to $\alpha\gamma$. This may be realized in practice by changing the duration and strength of the pulses that implement the desired 
operations. Alternatively one may employ unitary folding, where operations are added that do not affect the Hamiltonian evolution but increase the noise, e.g. for a unitary gate $\mathcal{u}$ one may use $\mathcal{U} \rightarrow \mathcal{U} \mathcal{U} \mathcal{U}^\dagger$~\cite{temmeErrorMitigationShortDepth2017, kandalaErrorMitigationExtends2019,endoHybridQuantumClassicalAlgorithms2021a,9259940, laroseMitiqSoftwarePackage2022}(see the Supplemental Materia).

For extrapolation to zero noise value, $\gamma=0$, we employ the Richardson extrapolation technique~\cite{burdenNumericalAnalysis2010, 9259940}. Given a sequence of noise parameters $\alpha_i \gamma$ with $1 = \alpha_0 < \alpha_1<...<\alpha_n$, the estimated noiseless expectation value for an operator $M$ is $ \langle M_E \rangle (\gamma = 0) = \sum_{k=0}^n \beta_k \langle M \rangle (\alpha_k \gamma)$. The coefficients  $\beta_k$ satisfy the relations $\sum^n_{k=0} \beta_k=1$ and $\sum^n_{l=0}\beta_l\alpha_l^k=0$, giving $\beta_k = \prod_{i \neq k}\frac{\alpha_i}{\alpha_i-\alpha_k}$~\cite{endoHybridQuantumClassicalAlgorithms2021a, temmeErrorMitigationShortDepth2017, kandalaErrorMitigationExtends2019}.


\paragraph{Matrix Product Density Operators.--} 
We use MPDOs to simulate noisy quantum circuits. MPDOs are an extension of matrix product states for simulating thermal or dissipative systems~\cite{verstraeteMatrixProductDensity2004a, zwolakMixedStateDynamicsOneDimensional2004a, SCHOLLWOCK201196}. 
We employ an approach based on the work of Zwolak and Vidal~\cite{zwolakMixedStateDynamicsOneDimensional2004a}.
Based on Choi's isomorphism $\ket{\psi}\bra{\phi} = \ket{\psi} \ket{\phi} = \vket{\psi \otimes \phi}$, one can vectorize the density matrix $\rho = \ket{\psi}  \bra{\psi} \rightarrow \vket{\rho} = \vket{\psi \otimes \psi} $. 
Similar to the matrix-product states representation for a quantum state~\cite{montangeroIntroductionTensorNetwork2018,javanmardSharpEntanglementThresholds2018,SCHOLLWOCK201196, ciracMatrixProductStates2021} $\ket{\psi} = \sum_{\{\sigma\}}{T^{\sigma_1}_{1} \ldots T^{\sigma_N}_{N} \ket{\sigma_1 \ldots \sigma_N}}$, 
we have in the MPDO representation $\vket{\rho} = \sum_{\{\sigma \sigma'\}}{M^{\sigma_1 \sigma'_1}_{1} \ldots M^{\sigma_N \sigma'_N}_{N} \ket{\sigma_1 \ldots \sigma_N}\bra{\sigma'_1 \ldots \sigma'_N}}$, with $\sigma_j$ possible single-site states, and  $M^{\sigma_i, \sigma'_i}_i = T^{\sigma_i}_i \otimes (T^{\sigma'_i}_i)^\star$. 
By reformulating the Liouvillian in a vectorized form $\mathcal{L}_{\#}$, and for a time-independent $\mathcal{L}_{\#}$, the time evolution 
is given by $\vket{\rho(T)} = e^{T\mathcal{L}_{\#}}\vket{\rho(0)}$. 
Here we consider the case where $\mathcal{L}_{\#}$ is a sum of nearest-neighbor interaction terms $\mathcal{L}_{\#}[\rho] = \sum_{\langle i,j \rangle} \mathcal{L}^{[i,j]}_{\#}[\rho]$. 
This structure allows for using time-evolving block decimation to simulate the time evolution~(see the Supplemental Material). 

In this work, we consider a chain with $N=32$ spins initially prepared in the state 
$\ket{\psi_0} = \ket{+}^{\bigotimes N}$, with $|+\rangle = H|\downarrow \rangle=\left ( |\downarrow\rangle + |\uparrow\rangle \right )/\sqrt{2}$, with $\{ |\uparrow\rangle,  
|\downarrow\rangle \}$ the  eigenstates of $\sigma^z$, and $H$ the Hadamard gate. 
We evolve this state using Hamiltonian ~\eqref{eq: TFI}. In our MPDO simulation, we use second-order Suzuki-Trotter decomposition~\cite{SCHOLLWOCK201196, hatanoFindingExponentialProduct2005} with time step $\delta t=0.01/J$. We discard states with Schmidt's values less than $10^{-5}$ during the algorithm run with maximum bond dimension $\chi_{max}=200$. Although one can reach longer times, we stopped at $t_{max}=14/J$.

\begin{figure}[t!]
\centering
\includegraphics[width=\columnwidth]{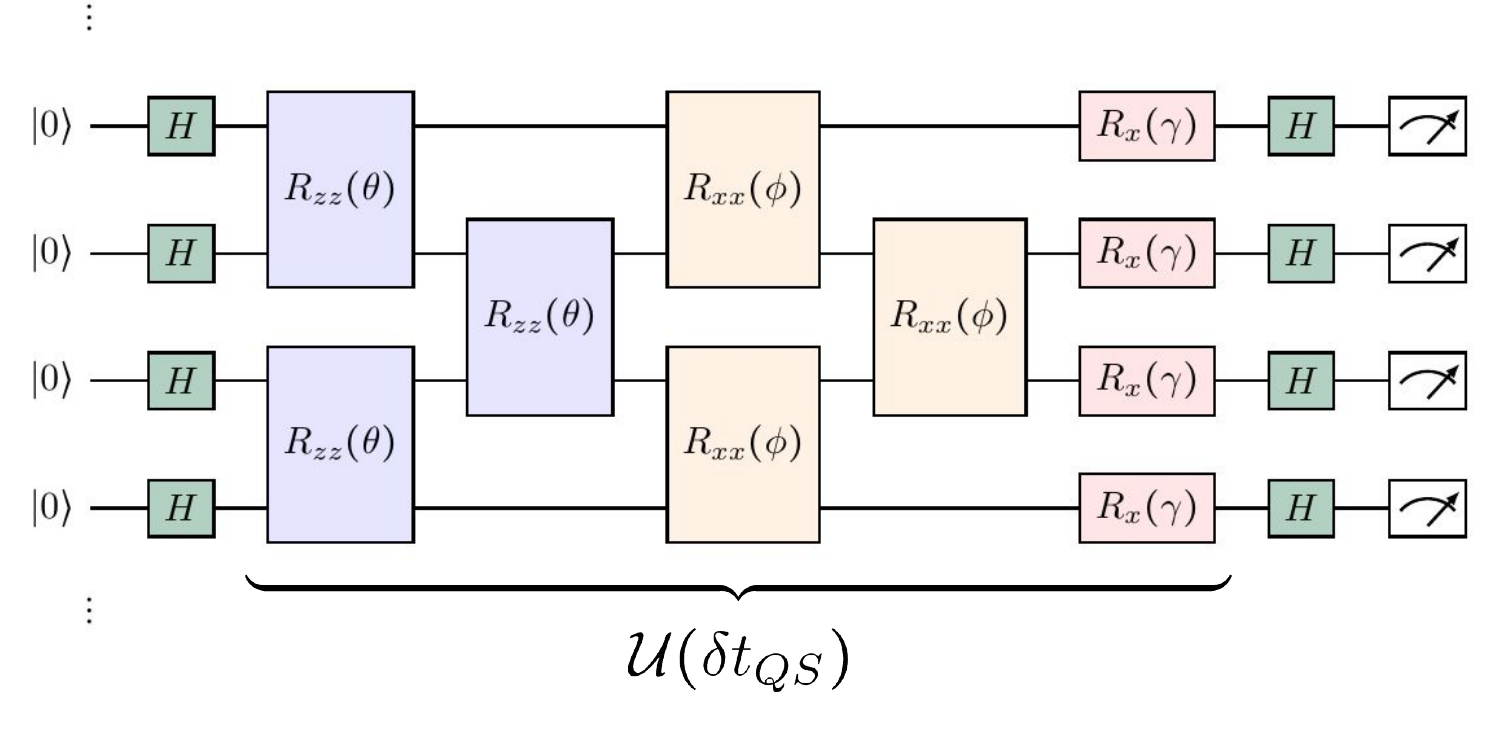}
\caption{Quantum circuit employed for the quantum simulation. The gate parameters are given by $\theta = -2\delta t_{QS} J_z$, $\phi = -2\delta t_{QS} J_x $, $\gamma = -2\delta t_{QS} h_x$. The part implementing the evolution operator $\mathcal{U}(\delta t_{QS})$ is repeated $M$ times to reach the desired time $t$~(see text).}
\label{fig1: qcircuit}
\end{figure}

\paragraph{Quantum simulation.--}
In addition to our MPDO calculations, we have evaluated the noisy dynamics using an IBMQ quantum simulator, which employs either the depolarizing noise model with the one-qubit probabilistic noise parameter $p=0.001$~(corresponding in our MPDO calculations to $\gamma=0.025 J$), or a noise model simulating device noise. The latter was simulated using the qiskit Aer noise package for the 7-qubit IBMQ Nairobi device~\cite{Qiskit}.
The time evolution given by the unitary operator $\mathcal{U}(t) = e^{-i \mathcal{H} t}$ is evaluated using Suzuki-Trotter decomposition into $M$ time-steps $\delta t_{QS}=t/M$, 
$\mathcal{U}(t) = [\mathcal{U}(\delta t_{QS})]^M$,  with $\mathcal{U}(\delta t_{QS}) = e^{-i \delta t_{QS} \mathcal{H}_{zz} } e^{-i\delta t_{QS} \mathcal{H}_{xx}}e^{-i\delta t_{QS} \mathcal{H}_{x}} + \mathcal{O}(\delta t^2_{QS})$~\cite{Smith2019, fausewehDigitalQuantumSimulation2021b}, with ${\cal H}_{xx}$~(${\cal H}_{zz}$) the $x$~($z$) spin-spin interactions, and ${\cal H}_x$ the part of the Hamiltonian given by the transversal field. 
The Loschmidt amplitude $\langle \psi_0| \mathcal{U}(t)|\psi_0 \rangle = \langle 0^{\otimes N}|H^{\otimes N} \mathcal{U}(t) H^{\otimes N} |0^{\otimes N}\rangle$ can be evaluated using a gate-based quantum computer with the circuit of Fig.~\ref{fig1: qcircuit}. \note{More precisely, the Loschmidt echo is extracted from the probability of the all-zero bitstring after the final Hadamard rotation,
$
P_{0\ldots 0}(t)=\left|\langle 0^{\otimes N}|H^{\otimes N}\mathcal{U}(t)H^{\otimes N}|0^{\otimes N}\rangle\right|^2
=\left|\langle \psi_0|\mathcal{U}(t)|\psi_0\rangle\right|^2=\Lambda(t),
$
and not from the expectation value of $Z^{\otimes N}$.} We use the mitiq library~\cite{laroseMitiqSoftwarePackage2022} in order to scale the quantum circuits for the ZNE task.




\begin{figure}[t!]
\includegraphics[width=1.0\columnwidth]{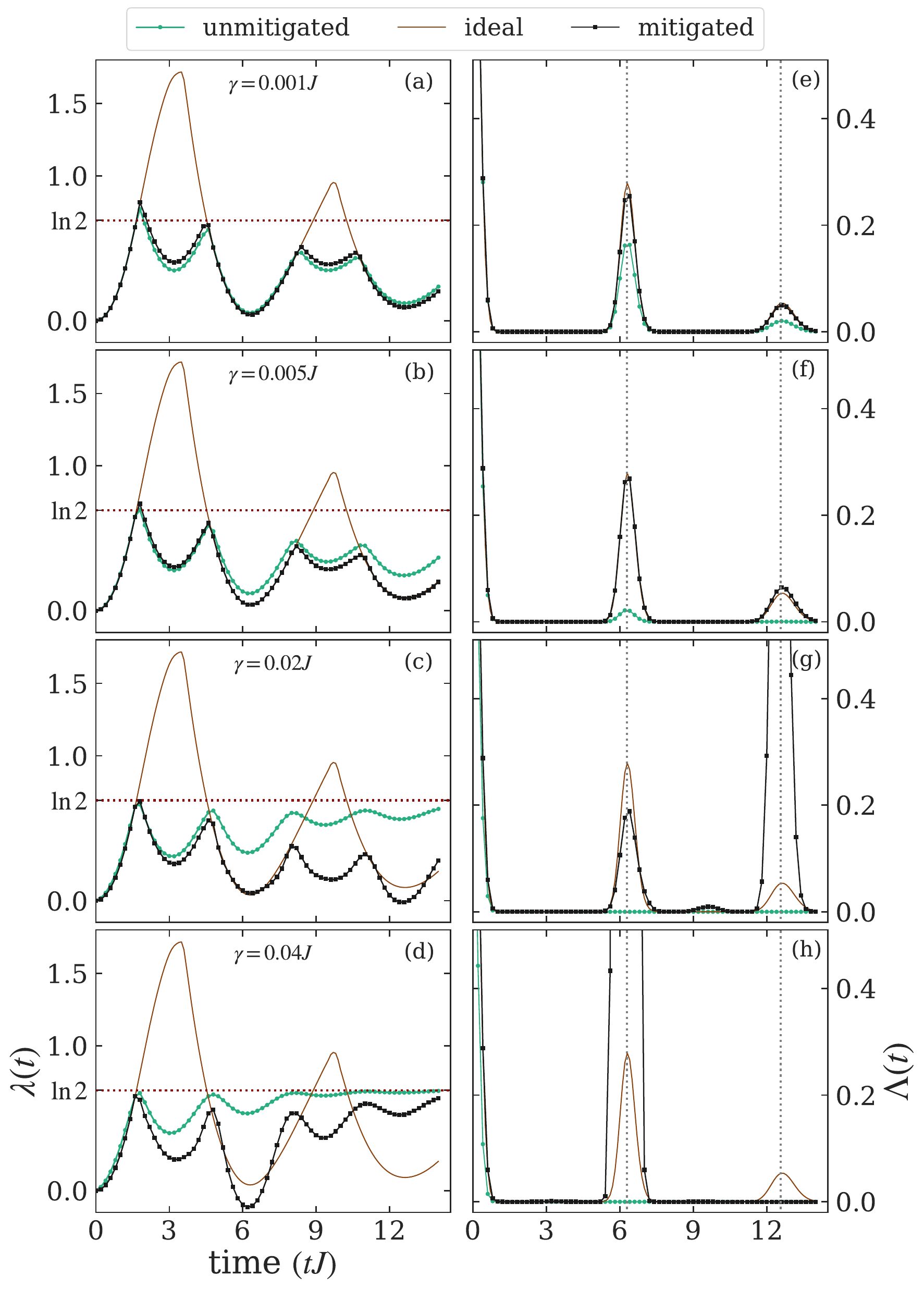}
\caption{Rate function $\lambda(t)$ as a function of time for different depolarization noise levels, $\gamma$. We show the results for different re-scaling factors $\alpha$, as well as the 
extrapolated value to zero noise. Loschmidt echo $\Lambda(t)= e^{-N\lambda(t)}$ for depolarizing noise for different noise strength $\gamma$. We show the results for the ideal value and the extrapolated value to zero noise.
For long-enough times and large-enough noise, the Loschmidt echo is bounded by a fully mixed state, which results in $\lambda = \ln{2}$, indicated by the red dashed line in each subplot. }
\label{fig2: DQPT-mpdo}
\end{figure}





\begin{figure}[t!]
\includegraphics[width=\columnwidth]{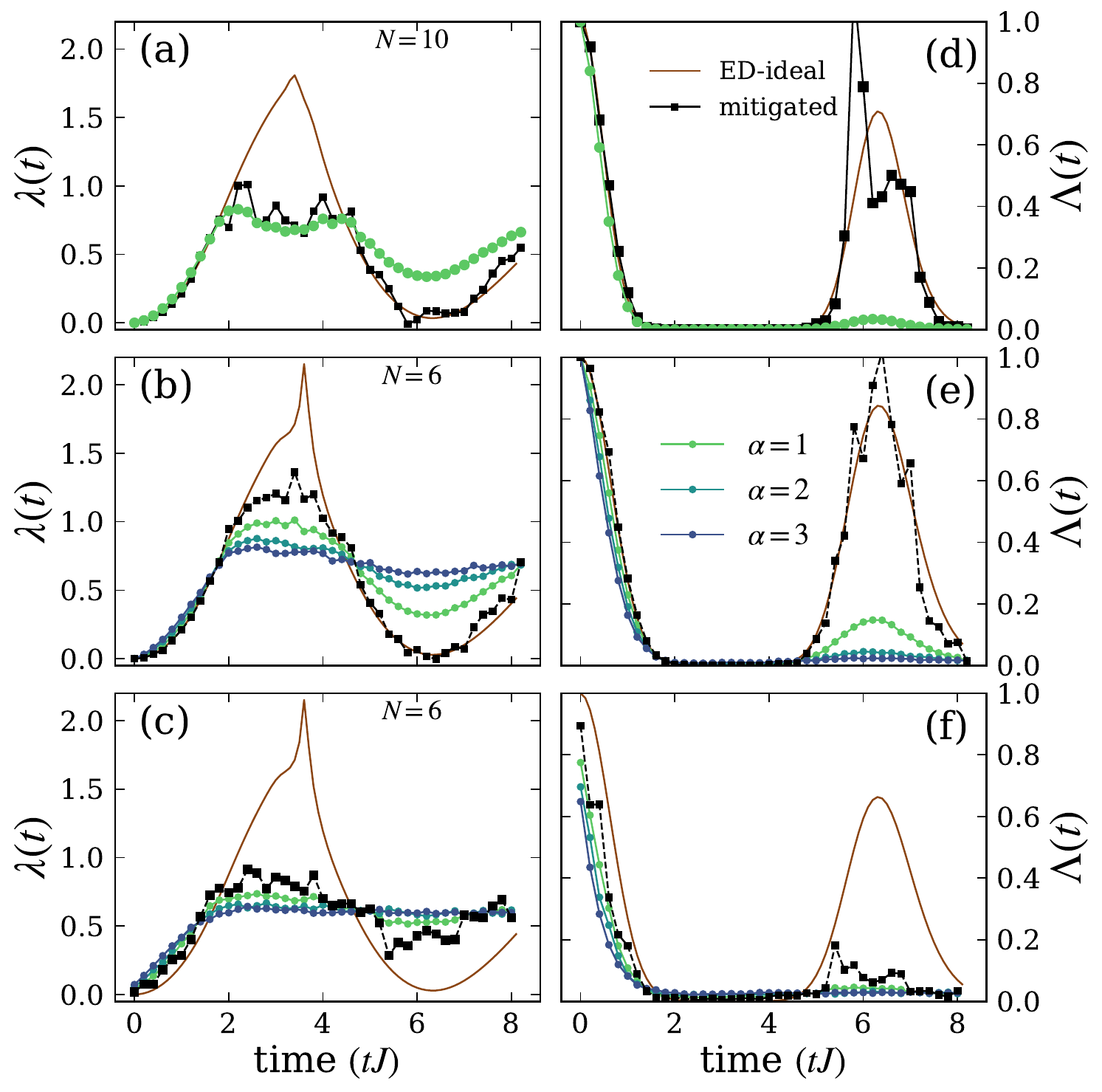}
\caption{Quantum simulation results of the behavior of the rate function 
$\lambda(t)$ and the Loschmidt echo $\Lambda(t)$. 
(a,d) and (b,e) show the results using the depolarizing noise model with one-qubit error $p=0.001$, for $N=10$ and $N=6$ qubits, respectively. (c,f) show the simulation results with the noise model of the IBMQ Nairobi 7-qubit device. In all the figures we depict the exact diagonalization~(ED) results, the unmitigated evolution, as well as the mitigated curve.
In Figs. (b,e) and (c,f) we show as well the 
results for different noise scaling factors $\alpha$. \note{ The number of shots performed to obtain the results was $N_s=16000$, which results in a negligible shot noise.}}
\label{fig3: QS-depolarizing}
\end{figure}




\begin{figure}[t!]
\includegraphics[width=\columnwidth]{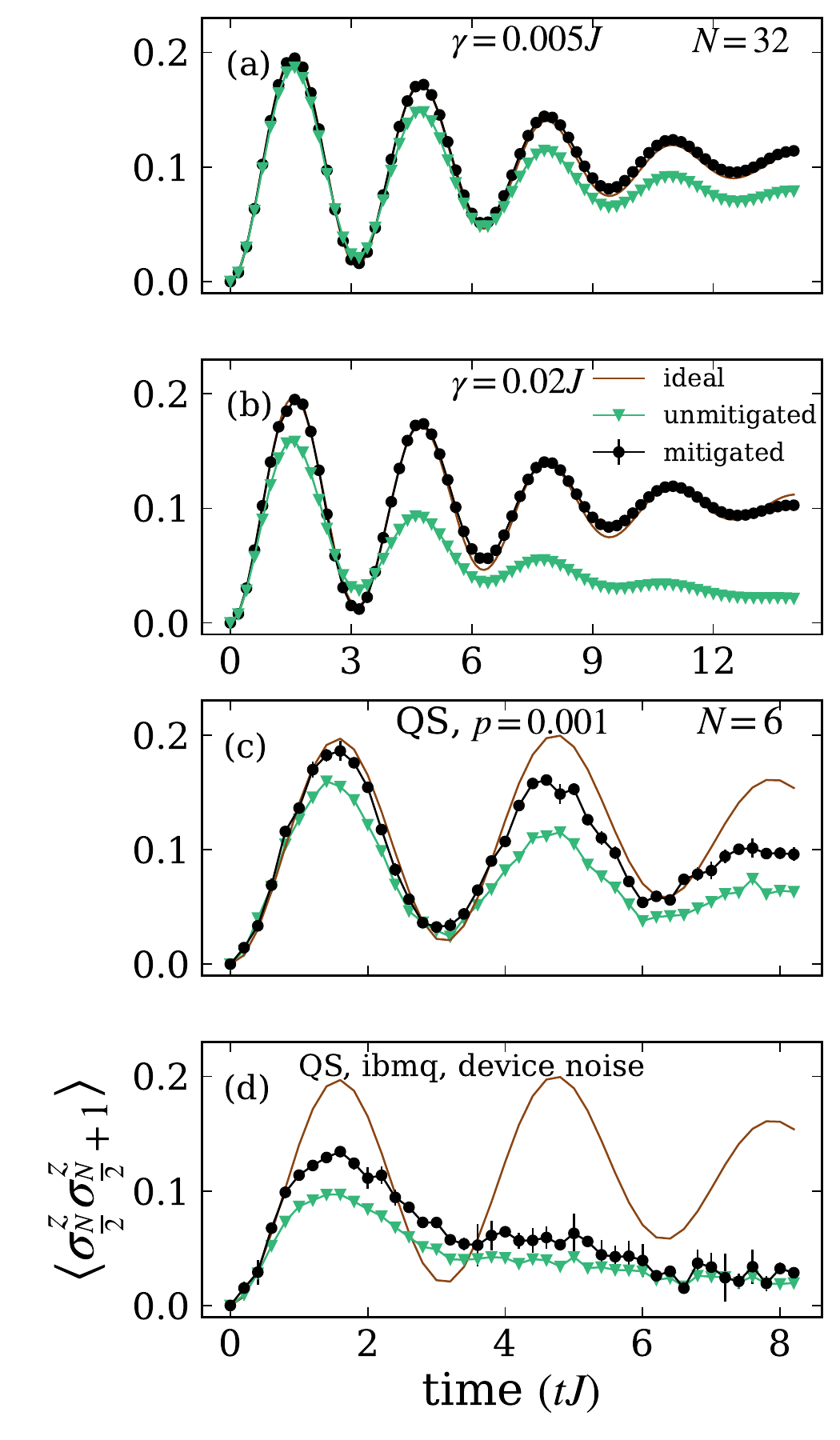}
\caption{Two-point correlation $C_{zz}(t)=\langle \sigma_{\frac{N}{2}}^z \sigma_{\frac{N}{2}+1}^z \rangle $. (a, b) show MPDO simulations results for depolarizing noise with noise strength $\gamma$ for $N=32$. (c, d) shows the quantum simulator results for $N=6$ for $\alpha=1.0, 1.5, 2.0$. (c) shows the results for depolarizing noise channel with $p=0.001$ and (d) shows the results using the device noise of the IBMQ Nairobi. We show the results for the ideal, the unmitigated and the extrapolated value to zero noise. \note{ The number of shots performed to obtain the results was $N_s=100000$, which results in a negligible shot noise.}}
\label{fig4: MPDO-ZNE, correlation}
\end{figure}




\begin{figure}[t!]
\includegraphics[width=\columnwidth]{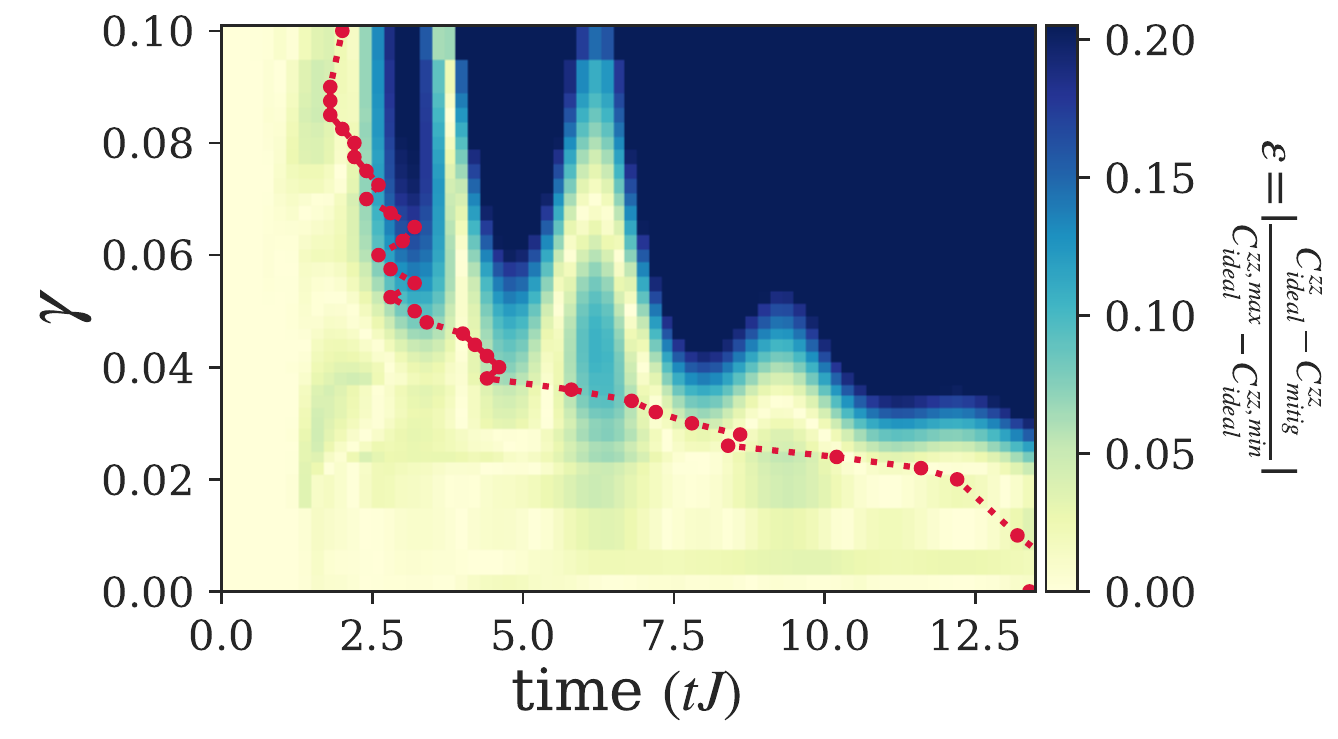}
\caption{Error rate $\varepsilon = \left|\frac{C^{zz}_{\mathrm{ideal}} - C^{zz}_{\mathrm{mitigated}}}{C^{zz, max}_{\mathrm{ideal}} - C^{zz, min}_{\mathrm{ideal}} } \right|$ as a function of noise strength $\gamma$ and time $tJ$. The red curve shows the threshold $\varepsilon<0.04$, below which the simulation faithfully retrieves the dynamics of Model~\eqref{eq: TFI}.}
\label{fig5: phase diagram}
\end{figure}


\paragraph{Loschmidt echo.--}
Figure~\ref{fig2: DQPT-mpdo} shows the return rate $\lambda(t)$ for different values of $\gamma$ versus the ideal (noise-free) case. \note{The extrapolation is performed on the return rate.}
For a low noise $\gamma \leq 0.01 J$, extrapolation recovers very well the ideal results, away from the DQPTs, all the way up to the maximal time considered $t=14/J$. 
For larger noise, ZNE may be significantly successful below a given noise and evolution-time threshold. In the right panels of Fig.~\ref{fig2: DQPT-mpdo} we show the Loschmidt echo $\Lambda(t)$. Note that ZNE recovers the first revival up to $\gamma=0.02 J$, although for an unmitigated noise the revival is basically lost for $\gamma\gtrsim 0.005 J$. For longer times and/or larger $\gamma$, ZNE fails eventually to recover, even qualitatively, the noise-free results. When accumulating noise, the system reaches a fully mixed state, $\rho_s(t) \equiv \frac{I}{2^N}$, such that $\Tr(\rho_0 \rho_s) \equiv \frac{1}{2^{N}}$ and the return rate is equivalent to $\lambda_s \rightarrow \ln{2}$.

The results in the vicinity of the DQPTs are considerably more affected by the noise. The expected peaks in $\lambda(t)$ associated to the DQPTs are not recovered even for a very low noise. Interestingly, for any noise, we observe rather a minimum of $\lambda(t)$, and hence a spurious revival of the Loschmidt echo,  within the time window at which the DQPT occurs in the noise-free system. This minimum is associated with a noise-induced splitting of the DQPT feature into a double-peak structure of $\lambda(t)$, consistent with two nearby noise-induced peaks of the return rate. As a result, ZNE cannot recover (even qualitatively) the noise-free DQPT. This points to an  inherent limitation of ZNE when applied to 
the quantum simulation of many-body dynamics.
\note{A simple phenomenological picture can be obtained by expanding the noisy Loschmidt echo near a clean dynamical critical time $t_c$. In the absence of noise, the Fisher-zero crossing of the Loschmidt amplitude produces a nonanalyticity in the return rate at $t_c$~\cite{heylDynamicalQuantumPhase2013, heylDynamicalQuantumPhase2018, Andraschko2014Loschmidt}. For a finite system, this local behavior may be represented by
\begin{equation}
\Lambda_0(t)\simeq A(t-t_c)^2,
\end{equation}
up to higher-order corrections. Consequently, the return rate $\lambda_0(t)=-(1/N)\log\Lambda_0(t)$ develops a sharp peak near $t_c$.

Depolarizing noise first of all lifts this zero. At the level of an effective phenomenological description, the noisy state contains an incoherent component, $\rho_\gamma(t)\simeq[1-\eta_\gamma(t)]\,\ket{\psi(t)}\!\bra{\psi(t)}+\eta_\gamma(t)\,I/2^N$, which gives
\begin{equation}
\Lambda_\gamma(t)=\Tr[\rho_0\,\rho_\gamma(t)]
\simeq [1-\eta_\gamma(t)]\,\Lambda_0(t)+\eta_\gamma(t)\,2^{-N}.
\label{eq:noisy-echo-floor-main}
\end{equation}
This incoherent floor regularizes the clean DQPT singularity: the Loschmidt echo no longer vanishes at $t_c$. By itself, however, such a floor would only round the single clean peak of the return rate. The observed double-peak structure requires an additional local effect of the noisy open-system dynamics: the curvature of the Loschmidt echo near the clean critical time is renormalized by the dissipative evolution. This is consistent with the broader observation that mixed-state and open-system settings can modify, shift, or generate dynamical critical structures~\cite{Bhattacharya2017MixedDQPT, Lang2018OpenDQPT}.

Writing $x=t-t_c$, we expand the smooth noisy Loschmidt echo locally around the clean critical time. Retaining the leading even terms of the approximately symmetric local profile gives the phenomenological Taylor/Landau-type form, with $b_\gamma>0$,
\begin{multline}
\Lambda_\gamma(t)\simeq
\Lambda_{\rm floor}(\gamma)
+a_\gamma\,(t-t_c)^2 \\
+b_\gamma\,(t-t_c)^4
+\mathcal{O}\!\left((t-t_c)^6\right).
\label{eq:landau-local-main}
\end{multline}
This expansion should not be interpreted as a microscopic derivation of the noisy peak positions. Rather, it is the minimal even local expansion that captures the conversion of a single minimum of the Loschmidt echo into two neighboring minima. For $a_\gamma>0$, $\Lambda_\gamma(t)$ has a single local minimum near $t_c$, corresponding to a single peak of the return rate. In the noisy DQPT window, however, the effective curvature can become negative, $a_\gamma<0$. Then $t_c$ becomes a local maximum of the Loschmidt echo, while two nearby local minima appear at
\begin{equation}
t_\pm=t_c\pm\sqrt{-\frac{a_\gamma}{2b_\gamma}}.
\label{eq:t-pm-main}
\end{equation}
Because the return rate is the negative logarithm of the Loschmidt echo, these two minima of $\Lambda_\gamma(t)$ appear as two smooth peaks of $\lambda_\gamma(t)$, while the region around $t_c$ appears as a local minimum of the return rate, i.e.\ as a spurious noise-induced revival of the Loschmidt echo.

This argument is not intended to predict the precise locations of the two peaks, which depend on the microscopic Liouvillian, the system size, and the full time dependence of the noisy state. Rather, it explains why weak depolarizing noise can transform a single clean DQPT peak into a pair of nearby noisy peaks: the incoherent component lifts the Fisher-zero singularity, while the noise-renormalized local curvature generates two neighboring minima of the Loschmidt echo. A more detailed phenomenological treatment is given in Supplemental Sec.~\ref{sec:supp_local_expansion}.}

\begin{figure}[h!]
\centering
\includegraphics[width=\columnwidth]{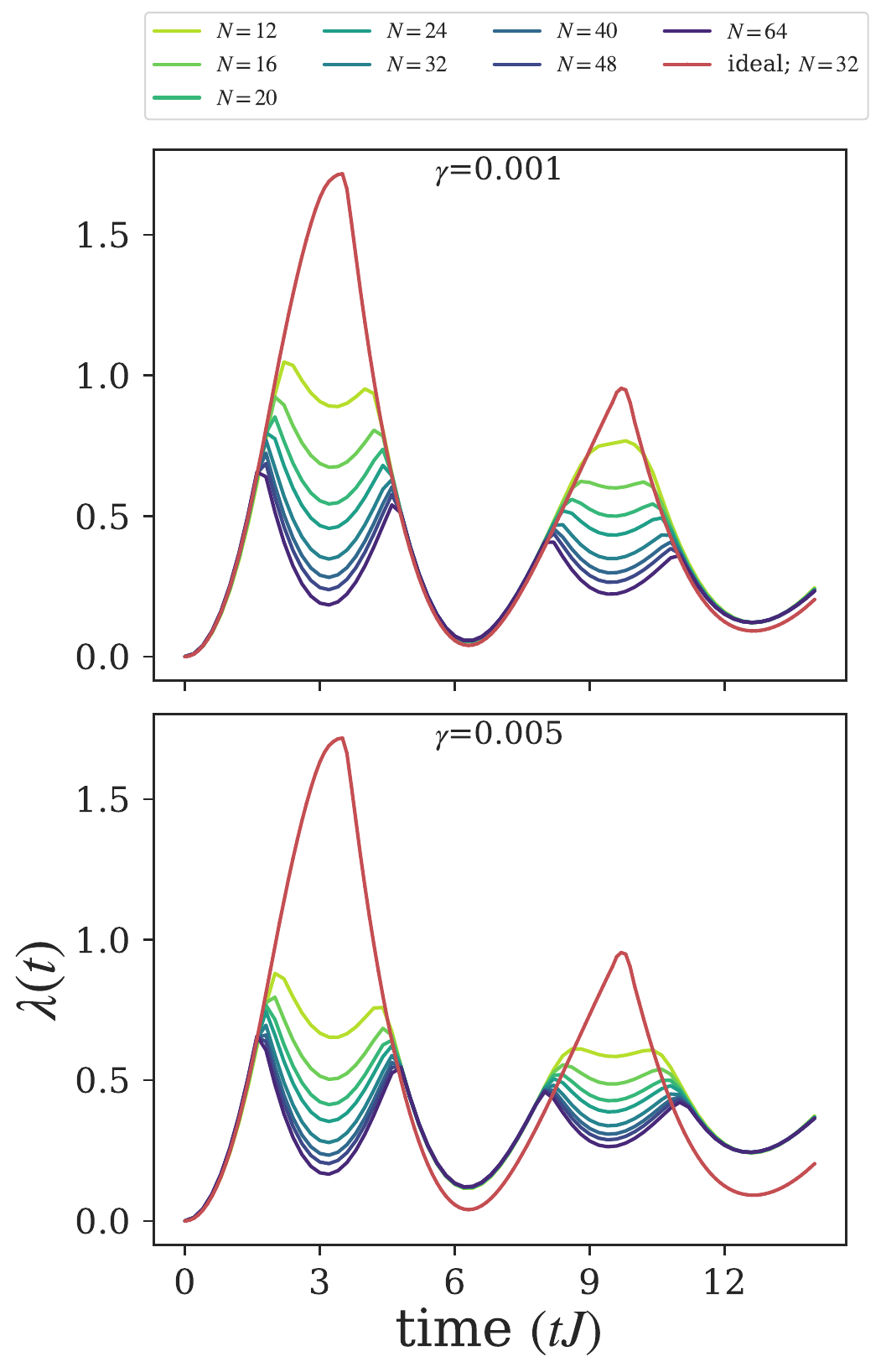}
\caption{System-size dependence of the return rate $\lambda(t)$ in the DQPT region for weak depolarizing noise. Results are shown for $\gamma=0.001J$ and $\gamma=0.005J$, with system sizes from $N=12$ to $N=64$. The ideal $N=32$ curve is shown for comparison. The noise-induced double-peak structure and the associated spurious revival near the clean critical time persist at very small noise strength and become more pronounced with increasing system size, indicating that the effect is not a finite-size or finite-shot artifact.}
\label{fig:size_dqpt}
\end{figure}

\note{The double-peak structure shown in Fig.~\ref{fig2: DQPT-mpdo} and its system-size dependence in Fig.~\ref{fig:size_dqpt} support the interpretation above. The MPDO curves are deterministic expectation values of the Lindblad open-system dynamics and therefore do not contain sampling noise. Moreover, the double-peak structure is already visible for very small depolarizing strengths and becomes more pronounced as $N$ is increased. Hence the observed splitting is not a finite-shot fluctuation or a small-system artifact, but a systematic feature of the noisy open-system dynamics.}

Figure~\ref{fig3: QS-depolarizing} shows our results obtained from 
the quantum simulator. For the case of depolarizing noise the results 
resemble those on the MPDO calculations~(see Figs.~\ref{fig3: QS-depolarizing}~(a,d) and~(b,e)). Note in particular that, as already observed in our MPDO calculations, using ZNE allows for the basically perfect recovery of the first revival of the Loschmidt echo, which is almost unobservable for the unmitigated case. 
Our quantum simulation results for up to $10$ spins, although significantly noisy in the DQPT region, suggest as well the
double-peak structure mentioned above~(see Fig.~\ref{fig3: QS-depolarizing}(a)). The MPDO study of systems with variable number of spins indicates that the double-peak structure, and the associated noise-induced spurious revival, become more prominent for a larger number of
spins. Hence, finite-size effects may explain the differences with the MPDO results in the DQPT window. The corresponding system-size dependence of the DQPT peak splitting is shown up to $N=64$ in Fig.~\ref{fig:size_dqpt}.

In contrast, the device-noise simulations, see Figs.~\ref{fig3: QS-depolarizing}~(c,f), show that the noise strength in the considered device is large, and the results approach the fully-mixed state. As a result, ZNE can faithfully recover the exact results only for very short times~($t<2/J$), while failing at later times. 


\paragraph{Two-site correlation.--}
The analysis of the two-site correlation function in the middle of the spin chain, $C^{zz}(t)=\langle \sigma_{\frac{N}{2}}^z \sigma_{\frac{N}{2}+1}^z \rangle$,   
provides further information about the ZNE performance. The zero-noise evolution is characterized by oscillations of period $\pi/J$, which damp to a finite value.
Figure~\ref{fig4: MPDO-ZNE, correlation} shows $C^{zz}(t)$ for different values of $\gamma$. We depict for each case the noise-free result 
$C_{\mathrm{ideal}}^{zz}(t)$, as well as the mitigated value 
$C_{\mathrm{mitigated}}^{zz}(t)$. For short times~(shallow circuits), $Jt<2$, ZNE is successful for all $\gamma$ up to $0.1 J$. For small values of the noise strength, $\gamma \leq 0.02 J$,  ZNE can recover the noise-free results up to the longest time considered. Note that, remarkably, this is so despite of the strong noise-induced damping observed  already for $\gamma=0.005 J$. For large $\gamma$ values noise results at long-enough times  in a vanishing correlation, rendering 
 the ZNE eventually unable to recover the noise-free result. 

In Fig.~\ref{fig5: phase diagram}, we depict the error rate, defined as $\varepsilon = \left|\frac{C^{zz}_{\mathrm{ideal}} - C^{zz}_{\mathrm{mitigated}}}{C^{zz, max}_{\mathrm{ideal}} - C^{zz, min}_{\mathrm{ideal}} } \right|$, 
as a function of both noise strength $\gamma$ and time $tJ$.  Our results show a threshold, given by the red curve in Fig.~\ref{fig5: phase diagram}, such that the ZNE is reliable below this threshold. In Fig.~\ref{fig5: phase diagram}, for a given $\gamma$ we have determined the threshold time as that in which $\varepsilon<0.04$. One can, however, 
employ other thresholds that would provide a similar picture.

Our quantum simulation results for $N=6$, using depolarizing noise, confirm the significant improvement of the $C_{zz}(t)$ calculation when using ZNE~(see Fig.~\ref{fig4: MPDO-ZNE, correlation}~(c)). In contrast, ZNE produces a more modest improvement for the device-noise calculation~(see Fig.~\ref{fig4: MPDO-ZNE, correlation}~(d)).



\paragraph{Conclusions.--}
We have investigated the performance of zero-noise extrapolation for the digital quantum simulation of the dynamics of many-body quantum systems. By means of matrix-product density operators, we have evaluated the particular case of the simulation of a quenched 
transverse-field Ising model in the presence of depolarizing noise. Our results 
show that zero-noise extrapolation allows to recover noise-free results, for both the Loschmidt echo and the inter-site correlation dynamics, below a given noise and evolution-time threshold. However, the method seems to be inherently unable to recover the ideal Loschmidt dynamics in the DQPT window, where depolarizing noise produces a robust double-peak structure and an associated spurious revival.
Our matrix-product density analysis, backed up by direct quantum simulations, provides a blueprint for the study of the performance of ZNE on noisy quantum processors with large number of qubits and deep circuits. Future work will address different noise models using the MPDO simulations for different quantum computing tasks, such as quantum optimization, and will benchmark these results with those on the actual device, developing a device-specific noise analysis. Another possibility would be to consider the role of cross-talks between qubits in the simulations and show how to mitigate them. Furthermore, one could expand such simulations to higher dimensions and a more realistic quantum chip geometry.

\acknowledgements
\paragraph{Acknowledgements.---}
We acknowledge funding by the Volkswagen foundation and the Ministry of Science and Culture of Lower Saxony through \emph{Quantum Valley Lower Saxony Q1 (QVLS-Q1)}, and 
by the Deutsche Forschungsgemeinschaft (DFG, German Research Foundation) -- under Germany's Excellence Strategy -- EXC-2123 Quantum-Frontiers -- 390837967.

\bibliography{references}
\bibliographystyle{apsrev4-1}

\clearpage 
\appendix
\onecolumngrid

\section*{\textemdash{} Supplemental Material \textemdash{}}
\section*{Quantum simulation of dynamical phase transitions in noisy quantum devices}

\begin{center}
Younes Javanmard, Ugne Liaubaite, Tobias J. Osborne, Luis Santos
\emph{Institut f\"ur Theoretische Physik, Leibniz Universit{\"a}t Hannover, Appelstraße 2, 30167 Hannover, Germany}
\end{center}

\vspace{4 pt}

\begin{description}
\item [{Summary}] We provide additional details and numerical results supplementing the conclusions from the main text.

\end{description}

\setcounter{equation}{0}
\setcounter{figure}{0}
\setcounter{table}{0}
\setcounter{page}{1}
\makeatletter
\renewcommand{\theequation}{S\arabic{equation}}
\renewcommand{\thefigure}{S\arabic{figure}}
\renewcommand{\bibnumfmt}[1]{[S#1]}
\renewcommand{\citenumfont}[1]{S#1}
   
\section{Matrix product density operators}
The vectorized Liouvillian is given by 
\eq{
    \mathcal{L}_{\#} & \equiv -i(\mathcal{H} \otimes I - I \otimes \mathcal{H}^T) \nonumber \\ 
                &+ \sum_\mu {(L_\mu \otimes L^*_\mu - \frac{1}{2}L^{\dagger}_\mu L_\mu \otimes I -\frac{1}{2}I \otimes L^*_\mu L^T_\mu)}
    \label{eq: vectorize master}}

\begin{figure}[th!]
\includegraphics[width=9cm]{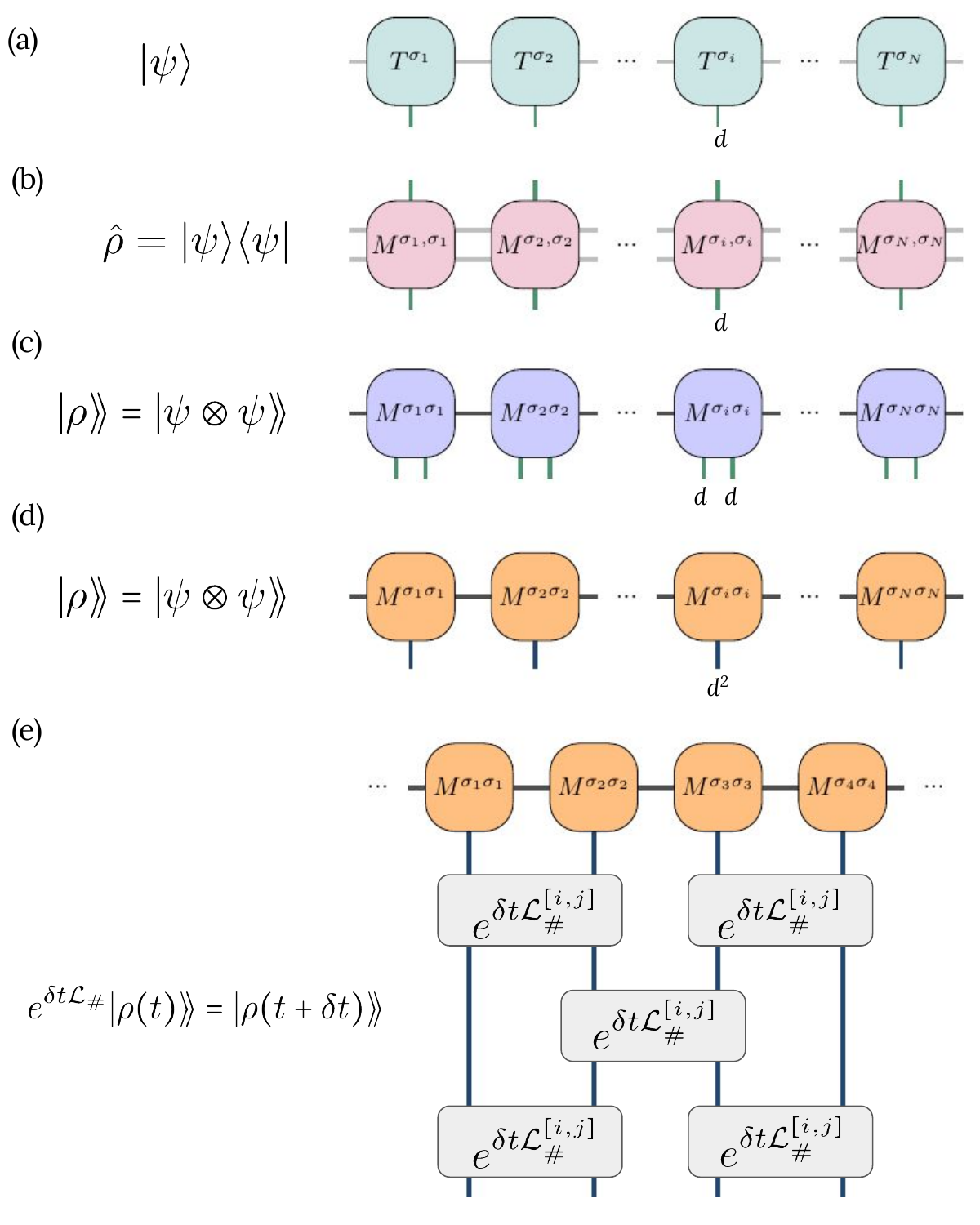}
\hspace{2cm}
\caption{MPDO steps to evolve a system based on the Liouvilian master equation.}
\label{fig1sup:mpdo}
\end{figure}

Fig.~\ref{fig1sup:mpdo} shows different steps for MPDO time evolution     
    
\begin{figure}[t!]
\includegraphics[width=0.6\columnwidth]{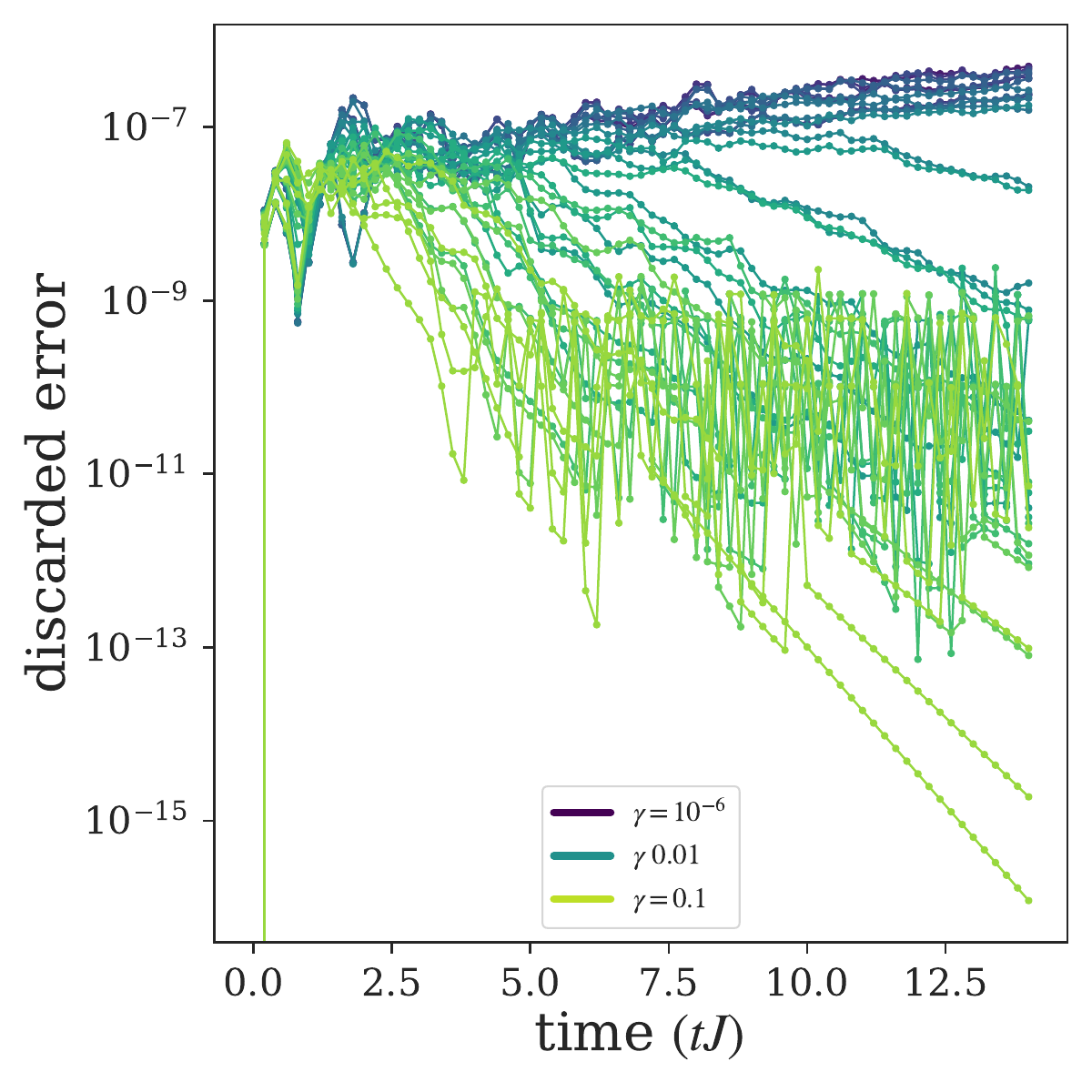}
\caption{Discarded error of the MPDO simulation for the main results in the paper. }
\label{fig2sup: DQPT-mpdo}
\end{figure}

\section{Depolarizing noise channel}
We comment in the following on the relation between the probabilistic error $p$ of the depolarizing channel, and the dissipation strength $\gamma$ in the Lindblad master equation: 
\eq{\Dot{\rho} 
=\gamma \sum_{\mu=x,y,z} \left ( -\rho + \sigma^\mu \rho \sigma^\mu \right ) 
= 2\gamma (I-2\rho).
\label{supeq: Lindblad eq}} 
Here we have employed the identity 
$\sum_\mu \sigma^\mu M \sigma^\mu = 2\Tr(M)I-M$, 
valid for any $2 \times 2$ matrix $M$, and $\Tr(\rho)=1$. 
Using the parametrization:
$\rho(t) = \frac{1}{2}(I + \Vec{r}(t)\cdot\vec{\sigma})$, 
we can rewrite the master equation in the form:
\eq{\dfrac{1}{2} \sum_\mu \Dot{r}_\mu\sigma^\mu  = -2\gamma \sum_\mu r_\mu \sigma^\mu,}
and hence
$r_\mu(t) = e^{-4 \gamma t} r_\mu(0)$. 
Then:
\eq{
\rho(\delta t)= \frac{1}{2}(I + e^{-4\gamma\delta t} \Vec{r}(0)\cdot\vec{\sigma}) = \left ( 1-p(\delta t) \right )\rho(0) + \frac{I}{2}p(\delta t),
}
which has the form of a depolarizing channel with  
$p(\delta t)=1-e^{-4\gamma\delta t}$.


\section{Benchmarking the MPDO simulations vs exact diagonalization}
\begin{figure}[h!]
\includegraphics[width=0.6\columnwidth]{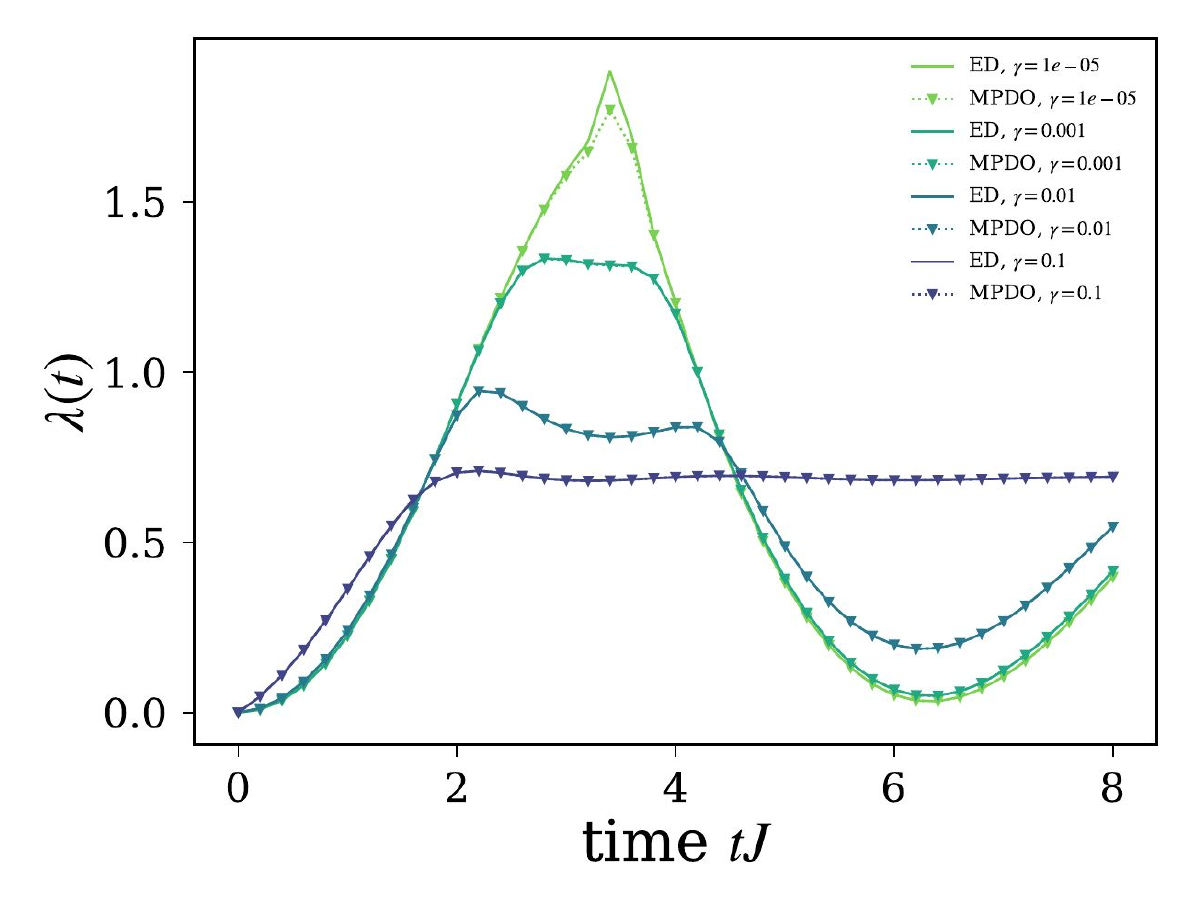}
\caption{Dynamics of model~\eqref{eq: Integrable TFIM} for $N=8$ sites.  
Comparison of the return rate $\lambda(t)$ evaluated with 
exact diagonalization~(ED) and with MPDO for different noise strengths. In our MPDO calculations we have employed a maximum bond dimension $\chi_{max}=200$.}
\label{fig2sup: EDvsMPDO}
\end{figure}

In order to benchmark our MPDO results, we compare in the following the results obtained using exact diagonalisation and those from the MPDO calculation. We consider a small system site ($N=8$) and different noise strengths.
We consider the integrable TFIM:
\eq{\mathcal{H} = \sum_i{ (\sigma_i^z \sigma_{i+1}^z + h \sigma_i^x)},
\label{eq: Integrable TFIM}
}
with $J = 1$ and $h=0.1$, and we quench from the same 
initial state as in the main text. Our results for the return rate  
depicted in Fig.~\ref{fig2sup: EDvsMPDO} show an excellent agreement between 
exact diagonalization and MPDO calculations. 
Note also that compared to our MPDO calculations with $N=32$ sites, we do not observe for $N=8$ the same behavior within the DQPT region, which rather resembles more closely our quantum simulation results. This indicates that the differences in the main text between the quantum simulation and MPDO results in the vicinity of the DQPT are due to the different system sizes.




\note{
\section{Check for fully mixed states}
Figure~\ref{fig5sup: EDmix} shows for different $N$ and $\gamma$ that the density matrix $\rho$ of the system tends to that of a fully-mixed state, characterized by $\Tr(\rho^2) =1/2^N$.

\begin{figure}[h!]
\includegraphics[width=0.6\columnwidth]{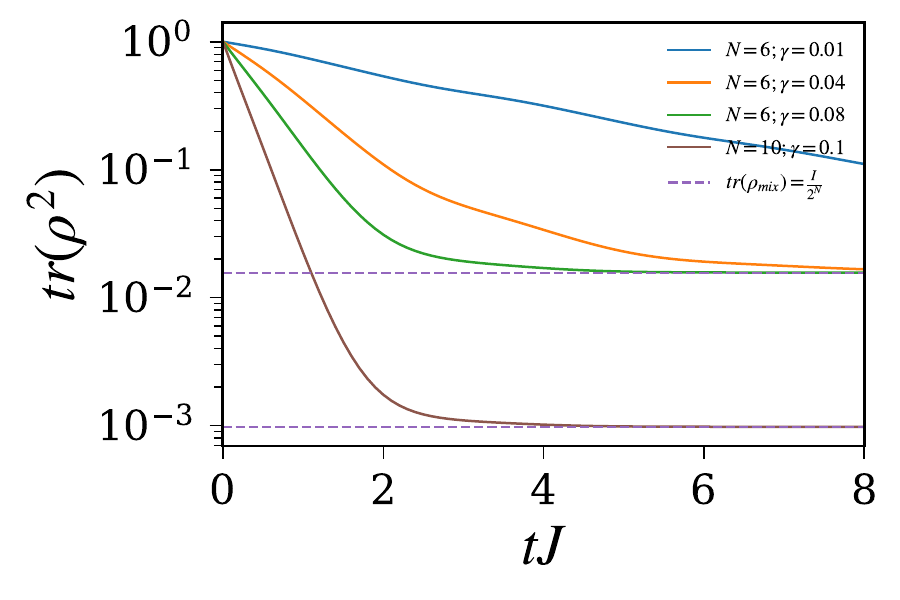}
\caption{The figure shows the behavior of time evolved state for different noise strength $\gamma$. As $\gamma$ increases, the state reaches a fully mixed state, i.e., $\Tr(\rho_{mix})=1/2^N$ (dashed lines). This justifies that $\lambda_s \rightarrow \ln{2}$. }
\label{fig5sup: EDmix}
\end{figure}

}
\note{\section{Depolarizing noise and local expansion near a DQPT}
\label{AppendixA}
\label{sec:supp-double-peak}
\label{sec:supp_local_expansion}

We first recall the clean behavior near a dynamical critical time. In the transverse-field Ising quench, DQPTs occur when Fisher zeros of the Loschmidt amplitude cross the real-time axis~\cite{heylDynamicalQuantumPhase2013, heylDynamicalQuantumPhase2018}. Close to such a critical time $t_c$, the relevant contribution to the Loschmidt amplitude has a simple zero. For the finite systems studied here, this local behavior may be represented phenomenologically by a quadratic zero of the Loschmidt echo,
\begin{equation}
\Lambda_0(t)\simeq A(t-t_c)^2.
\label{eq:clean-quadratic-zero}
\end{equation}
The corresponding return rate,
\begin{equation}
\lambda_0(t)=-\frac{1}{N}\log\Lambda_0(t),
\end{equation}
therefore develops a sharp peak near $t_c$.

The local single-qubit depolarizing channel used in the main text is
\begin{equation}
\mathcal{E}_i(\rho_i)=(1-p)\,\rho_i+p\,\frac{I}{2}.
\end{equation}
For the purpose of a simple phenomenological description of the Loschmidt echo near a DQPT, we approximate the accumulated effect of depolarization by an effective global incoherent admixture,
\begin{equation}
\rho_\gamma(t)\simeq[1-\eta_\gamma(t)]\,\ket{\psi(t)}\!\bra{\psi(t)}
+\eta_\gamma(t)\,\frac{I}{2^N}.
\label{eq:effective-global-admixture}
\end{equation}
This effective form should be distinguished from the microscopic local channel. It is used only to capture the incoherent background generated in the return probability. It gives
\begin{equation}
\Lambda_\gamma(t)
=\Tr[\rho_0\,\rho_\gamma(t)]
\simeq [1-\eta_\gamma(t)]\,\Lambda_0(t)+\eta_\gamma(t)\,2^{-N}.
\label{eq:noisy-echo-floor}
\end{equation}
Thus depolarization lifts the zero of the clean Loschmidt echo and regularizes the singularity of the clean return rate.

This lifting alone, however, does not explain the observed double-peak structure. A term of the form $A(t-t_c)^2+\Lambda_{\rm floor}$ would simply produce a rounded single peak in the return rate. The splitting observed in the MPDO data indicates that the dissipative dynamics also changes the local curvature of the noisy Loschmidt echo around the clean critical time. We therefore use the more general local expansion
\begin{equation}
\Lambda_\gamma(t)\simeq
\Lambda_{\rm floor}(\gamma)
+a_\gamma\,(t-t_c)^2
+b_\gamma\,(t-t_c)^4
+\mathcal{O}\!\left((t-t_c)^6\right),
\qquad b_\gamma>0.
\label{eq:landau-local}
\end{equation}
In the clean limit, $a_\gamma>0$, so the echo has a single local minimum at $t_c$. In the noisy DQPT window, the effective quadratic coefficient can become negative, $a_\gamma<0$. The stationary points of the local expansion satisfy
\begin{equation}
\frac{d\Lambda_\gamma}{dt}
=2a_\gamma(t-t_c)+4b_\gamma(t-t_c)^3=0,
\end{equation}
and hence
\begin{equation}
t=t_c,\qquad
t_\pm=t_c\pm\sqrt{-\frac{a_\gamma}{2b_\gamma}}.
\label{eq:t-pm-quartic}
\end{equation}
For $a_\gamma<0$, $t=t_c$ is a local maximum of the Loschmidt echo, while $t_\pm$ are two local minima. Since
\begin{equation}
\lambda_\gamma(t)=-\frac{1}{N}\log\Lambda_\gamma(t),
\end{equation}
the two minima of the Loschmidt echo correspond to two peaks of the return rate. The local maximum of $\Lambda_\gamma(t)$ at $t_c$ corresponds to the spurious noise-induced revival seen as a local minimum in $\lambda_\gamma(t)$.

This local expansion should be viewed as a phenomenological explanation of the mechanism, not as a microscopic prediction of the exact peak locations. The coefficients $\Lambda_{\rm floor}(\gamma)$, $a_\gamma$, and $b_\gamma$ depend on the full noisy Liouvillian, the system size, and the microscopic dynamics. The expansion nevertheless captures the two essential ingredients observed numerically: the incoherent depolarizing background removes the clean Fisher-zero singularity, while the noise-renormalized curvature can split the clean DQPT peak into two smooth noisy peaks.

Finally, the observed splitting is not attributable to shot noise or numerical fluctuations. The MPDO results are deterministic expectation values of the Lindblad dynamics and do not involve finite sampling. The double-peak structure appears already at very small depolarizing strengths and persists as the noise strength is varied. In addition, the finite-size analysis in Fig.~\ref{fig:size_dqpt} shows that the double-peak structure becomes more pronounced with increasing system size, supporting its interpretation as a genuine open-system effect of depolarizing noise near a DQPT.
}
\note{\section{Circuit error rate and accumulated depolarization after \(M\) Trotter steps}
For a Trotter step \(\delta t\), the local depolarizing probability is \(p_{\delta t}=1-e^{-4\gamma\delta t}\). For a circuit with \(M=t/\delta t\) Trotter steps on \(N\) qubits, the effective probability that at least one local depolarizing event occurs is
\[
p_{\mathrm{circ}}(t)=1-(1-p_{\delta t})^{NM}\simeq NM\,p_{\delta t}\simeq 4N\gamma t,
\]
valid for \(p_{\delta t}\ll 1\).

A simple phenomenological description of the Loschmidt echo after \(M\) Trotter steps is
\[
\Lambda_\gamma(t)\simeq e^{-k\gamma t}\Lambda(t)+\bigl(1-e^{-k\gamma t}\bigr)c,
\]
where \(k\) is the number of noisy local channels per Trotter step and \(c\) is an effective incoherent background, with \(c=2^{-N}\) in the fully mixed limit. Near a clean DQPT, where \(\Lambda(t_c)=0\), depolarization lifts the zero of the echo to a finite floor and therefore rounds the cusp of \(\lambda(t)\), providing the basis for the peak-splitting estimate discussed in the previous section.
}

\section{Noise scaling in the quantum simulation}
Noise scaling in gate-based quantum computation is based on unitary folding, which consists in replacing a unitary operation $U$ by $U \rightarrow U(U^\dagger U)^n$ with $n$ being a positive integer. As $U^\dagger U =I$, folding does not have any logical effect. However, the noise is increased with the prolongation of the circuit. 
The mitiq library is used to scale the quantum circuits for zero noise extrapolation. The scaling is done using folding at random introduced in \cite{laroseMitiqSoftwarePackage2022, 9259940}. Before scaling the circuits, one needs to transpile the quantum circuits using a set of universal basis gates, so that the depth of the circuit ran is precisely known; Here we have used basis gates$=[u1, u2, u3, cx]$.

The value of the scaling factor $\alpha$ is important for the performance of ZNE. There are two options for unitary folding, circuit-folding and layer- or gate-based folding. The methods differ in the values of scaling factor $\alpha$ that can be reached. In circuit-folding the entire circuit is folded, and the values of $\alpha$ are of form 
$\alpha = 1+2n$ where $n$ is the number of identities added. In order to get finer scaling, i.e. $\alpha$ being able to reach all real numbers, one can scale part of a circuit~(folding only a selected number of layers from the entire circuit). The values of $\alpha$ are then of form $\alpha = 1 +(2k)/d$ where $k$ is any natural number and $d$ is the depth of the circuit. In this work we have employed layer-based folding. For more detailed information see Ref.~\cite{9259940}.

\section{Shot noise and sampling overhead}
\label{sec:shot-noise}

The MPDO and exact Lindblad results shown in the main text are deterministic expectation values obtained from the evolved density matrix $\rho(t)$. They therefore do not contain finite-shot sampling noise. This distinction is important in the DQPT region, where the Loschmidt echo can become exponentially small.

In a gate-based measurement, the Loschmidt echo is obtained from the probability of the all-zero bitstring after the final basis rotation,
\begin{equation}
\Lambda(t)=P_{0\cdots 0}(t)
=\bra{0\cdots 0}\rho(t)\ket{0\cdots 0}.
\label{eq:Lambda-bornprob}
\end{equation}
A finite-shot estimate is therefore binomial,
\begin{equation}
\widehat{\Lambda}(t)=\frac{n_{0\cdots 0}}{N_{\rm shots}},
\qquad
\mathrm{Var}\!\left[\widehat{\Lambda}(t)\right]
=\frac{\Lambda(t)\,[1-\Lambda(t)]}{N_{\rm shots}}.
\end{equation}
For the return rate, $\lambda(t)=-(1/N)\log\Lambda(t)$, error propagation gives
\begin{equation}
\Delta\lambda(t)
\simeq
\frac{1}{N}\sqrt{
\frac{1-\Lambda(t)}{N_{\rm shots}\,\Lambda(t)}}.
\end{equation}
Thus, when $\Lambda(t)\ll 1$, resolving the return rate with fixed accuracy requires a number of shots scaling as
\begin{equation}
N_{\rm shots}\sim \frac{1}{\Lambda(t)}=e^{N\lambda(t)}.
\label{eq:shot-overhead}
\end{equation}
For example, for $N=32$ and $\lambda=1.5$, one has $\Lambda=e^{-48}\simeq 1.4\times 10^{-21}$, so a direct estimate of the all-zero probability would require on the order of $1/\Lambda\sim 7\times 10^{20}$ shots. Hence finite sampling would make the Loschmidt echo in the DQPT region extremely difficult to resolve experimentally. It cannot, however, explain the double-peak structure observed in the deterministic MPDO/Lindblad curves; that structure is already present at the level of the open-system dynamics.

For small systems, where the full diagonal probability vector $p_i(t)=\bra{i}\rho(t)\ket{i}$ can be stored explicitly, finite sampling can be mimicked by drawing bitstrings from this distribution using rejection sampling~\cite{robert2004monte}. For large systems, where $2^N$ enumeration is intractable, the same diagonal distribution can instead be sampled sequentially from conditional probabilities obtained by contracting the MPDO; this procedure is described in Sec.~\ref{app:perfect_sampling_mpdo}.

\note{\section{Perfect sampling of bitstrings from an MPDO}
\label{app:perfect_sampling_mpdo}

This appendix records the conditional perfect-sampling algorithm used to generate computational-basis bitstrings from a matrix product density operator (MPDO). The algorithm samples from the probability distribution
\[
p(z_1,\ldots,z_N)
=
\langle z_1\cdots z_N|\rho|z_1\cdots z_N\rangle,
\qquad
z_j\in\{0,1\},
\]
defined by the diagonal of the density operator \(\rho\) in the computational basis.

\subsection{Projector formulation of sequential sampling}

Let
\[
P_0=\ket{0}\!\bra{0},
\qquad
P_1=\ket{1}\!\bra{1}.
\]
The probability of a full computational-basis outcome
\(
z=(z_1,\dots,z_N)
\)
is
\begin{equation}
p(z_1,\dots,z_N)
=
\Tr\!\bigl[(P_{z_1}\otimes\cdots\otimes P_{z_N})\,\rho\bigr].
\label{eq:full_projector_prob}
\end{equation}
This distribution can be sampled sequentially using the chain rule,
\begin{equation}
p(z_1,\dots,z_N)
=
p(z_1)\prod_{j=2}^{N} p(z_j\mid z_1,\dots,z_{j-1}),
\label{eq:chain_rule_projector}
\end{equation}
where the first-site probability is
\begin{equation}
p(z_1)
=
\Tr\!\bigl[(P_{z_1}\otimes I^{\otimes N-1})\,\rho\bigr],
\label{eq:first_site_prob}
\end{equation}
and, for \(j\ge 2\),
\begin{equation}
p(z_j\mid z_1,\dots,z_{j-1})
=
\frac{
\Tr\!\bigl[(P_{z_1}\otimes\cdots\otimes P_{z_j}\otimes I^{\otimes N-j})\,\rho\bigr]
}{
\Tr\!\bigl[(P_{z_1}\otimes\cdots\otimes P_{z_{j-1}}\otimes I^{\otimes N-j+1})\,\rho\bigr]
}.
\label{eq:conditional_projector_prob}
\end{equation}
Thus, once the prefix \(z_1,\dots,z_{j-1}\) has been sampled, one computes the two conditional probabilities corresponding to the local projectors \(P_0\) and \(P_1\) at site \(j\), draws the next bit \(z_j\), and proceeds to the next site.

\subsection{MPDO representation and diagonal blocks}

We write the MPDO in open-boundary form as
\begin{equation}
\rho
=
\sum_{\substack{s_1,\dots,s_N\\ s'_1,\dots,s'_N}}
M^{[1]\,s_1,s'_1}
M^{[2]\,s_2,s'_2}\cdots
M^{[N]\,s_N,s'_N}
\;
|s_1\cdots s_N\rangle\langle s'_1\cdots s'_N|,
\label{eq:mpdo_open_boundary}
\end{equation}
where each \(M^{[j]\,s_j,s'_j}\) is a matrix on the virtual bond space of dimension \(D_{j-1}\times D_j\), with \(D_0=D_N=1\). We assume that \(\rho\) is normalized,
\begin{equation}
\Tr(\rho)=1.
\end{equation}

Since we sample computational-basis outcomes, only the diagonal physical blocks contribute. We therefore define
\begin{equation}
D^{[j]}(z_j):=M^{[j]\,z_j,z_j},
\qquad
z_j\in\{0,1\}.
\label{eq:diagonal_blocks}
\end{equation}
Then the full bitstring probability in Eq.~\eqref{eq:full_projector_prob} becomes
\begin{equation}
p(z_1,\dots,z_N)
=
D^{[1]}(z_1)D^{[2]}(z_2)\cdots D^{[N]}(z_N),
\label{eq:bitstring_prob_contracted}
\end{equation}
where the product denotes contraction over the virtual bonds from left to right. In this way, inserting the local projector \(P_{z_j}\) at site \(j\) simply selects the diagonal MPDO block \(D^{[j]}(z_j)\).

\subsection{Environment formulation of the conditional probabilities}

To evaluate the conditional probabilities efficiently, we introduce left and right environments. The left environment associated with the sampled prefix is
\begin{equation}
L_0:=1,
\qquad
L_j:=L_{j-1}D^{[j]}(z_j),
\qquad
j=1,\dots,N.
\label{eq:left_env_mpdo}
\end{equation}
Thus \(L_j\) is a row vector on the bond between sites \(j\) and \(j+1\), encoding the contraction of the sampled prefix.

The right environments are defined recursively by
\begin{equation}
R_{N+1}:=1,
\qquad
R_j:=\sum_{b\in\{0,1\}} D^{[j]}(b)\,R_{j+1},
\qquad
j=N,\dots,1.
\label{eq:right_env_mpdo}
\end{equation}
Each \(R_j\) is a column vector containing the total contribution of all suffixes from site \(j\) onward. In particular,
\begin{equation}
R_1=\Tr(\rho)=1.
\end{equation}

Given the previously sampled prefix \(z_{<j}=(z_1,\dots,z_{j-1})\), the unnormalized weight for choosing \(z_j=b\in\{0,1\}\) is
\begin{equation}
\widetilde{\pi}_j(b)
=
L_{j-1}\,D^{[j]}(b)\,R_{j+1}.
\label{eq:unnormalized_weight_mpdo}
\end{equation}
The normalized conditional probability is therefore
\begin{equation}
\pi_j(b\mid z_{<j})
=
\frac{L_{j-1}\,D^{[j]}(b)\,R_{j+1}}
{\sum_{b'\in\{0,1\}} L_{j-1}\,D^{[j]}(b')\,R_{j+1}}.
\label{eq:conditional_prob_mpdo}
\end{equation}
This is exactly the MPDO implementation of the projector-based rule in Eq.~\eqref{eq:conditional_projector_prob}. Repeating this procedure from \(j=1\) to \(N\) produces a full bitstring distributed according to \(p(z_1,\dots,z_N)\). Since the sample is generated directly from the exact conditional probabilities, the method yields perfect samples without any Markov-chain equilibration.

\begin{algorithm}[H]
\caption{Conditional perfect sampling of bitstrings from an MPDO}
\label{alg:bitstring_sampling_mpdo}
\begin{algorithmic}[1]
    \STATE \textbf{input:} MPDO tensors $\{M^{[j]\,s,s'}\}_{j=1}^{N}$ and number of samples $\mathcal{N}_{\mathrm{samp}}$.
    \STATE \textbf{output:} Samples $\{z^{(\mu)}\}_{\mu=1}^{\mathcal{N}_{\mathrm{samp}}}$, where $z^{(\mu)}\in\{0,1\}^{\otimes N}$.
    \item[]
    \FOR{$\mu=1$ \TO $\mathcal{N}_{\mathrm{samp}}$}
        \STATE Set $L=1$ and initialize $z^{(\mu)}$ as an empty bitstring.
        \FOR{$j=1$ \TO $N$}
            \STATE Compute $\widetilde{\pi}_j(b)$ for $b\in\{0,1\}$, corresponding to inserting the local projector $P_b=\ket{b}\!\bra{b}$ at site $j$.
            \STATE Set $\pi_j(b)=\widetilde{\pi}_j(b)\big/\sum_{b'\in\{0,1\}}\widetilde{\pi}_j(b')$.
            \STATE Sample $z_j\sim \pi_j(\cdot)$ and append $z_j$ to $z^{(\mu)}$.
            \STATE Update $L$ using the diagonal block $M^{[j]\,z_j,z_j}$.
        \ENDFOR
    \ENDFOR
    \item[]
    \STATE \textbf{return:} $\{z^{(\mu)}\}_{\mu=1}^{\mathcal{N}_{\mathrm{samp}}}$.
\end{algorithmic}
\end{algorithm}

\subsection{Validation of perfect sampling: observables and Hamming-weight distributions}
\label{app:perfect_sampling_validation}

We demonstrate the perfect-sampling procedure of Alg.~\ref{alg:bitstring_sampling_mpdo} on the noisy non-integrable transverse-field Ising chain with $N=16$, bond dimension $\chi_{\max}=64$, and single-qubit depolarizing rate $\gamma=0.02$. The initial state is the fully $x$-polarized product state $\ket{+}^{\otimes N}$. At each snapshot time $t\in\{0.4, 0.8,\ldots,4.0\}$ we draw $N_{\rm shots}=8000$ perfect bitstring samples from $\rho(t)$ as computed by the MPDO TEBD evolution.

Fig.~\ref{fig:sup_perfect_sampling_czz} shows the centre-of-chain two-point correlator $C_{zz}(N/2, N/2{+}1)=\langle \sigma_z^{N/2}\sigma_z^{N/2+1}\rangle$ reconstructed from the bitstring samples as the empirical mean $(1/N_{\rm shots})\sum_\mu (1-2 b^{(\mu)}_{N/2})(1-2 b^{(\mu)}_{N/2+1})$, compared against the direct MPDO contraction. The sample estimator tracks the direct value within its statistical uncertainty, confirming that the conditional perfect-sampling procedure faithfully reproduces diagonal observables of $\rho(t)$.

Fig.~\ref{fig:sup_perfect_sampling_hamming_grid} reports the empirical Hamming-weight histograms $\hat p(k)=\#\{\mu:|b^{(\mu)}|=k\}/N_{\rm shots}$ at the ten snapshot times. The thin step line indicates the binomial $\mathrm{Binom}(N,1/2)$ distribution corresponding to the maximally-mixed Z-diagonal limit; the depolarizing dynamics drive the empirical histogram smoothly toward this reference over the simulated interval.

\begin{figure}[h!]
\centering
\includegraphics[width=0.85\columnwidth]{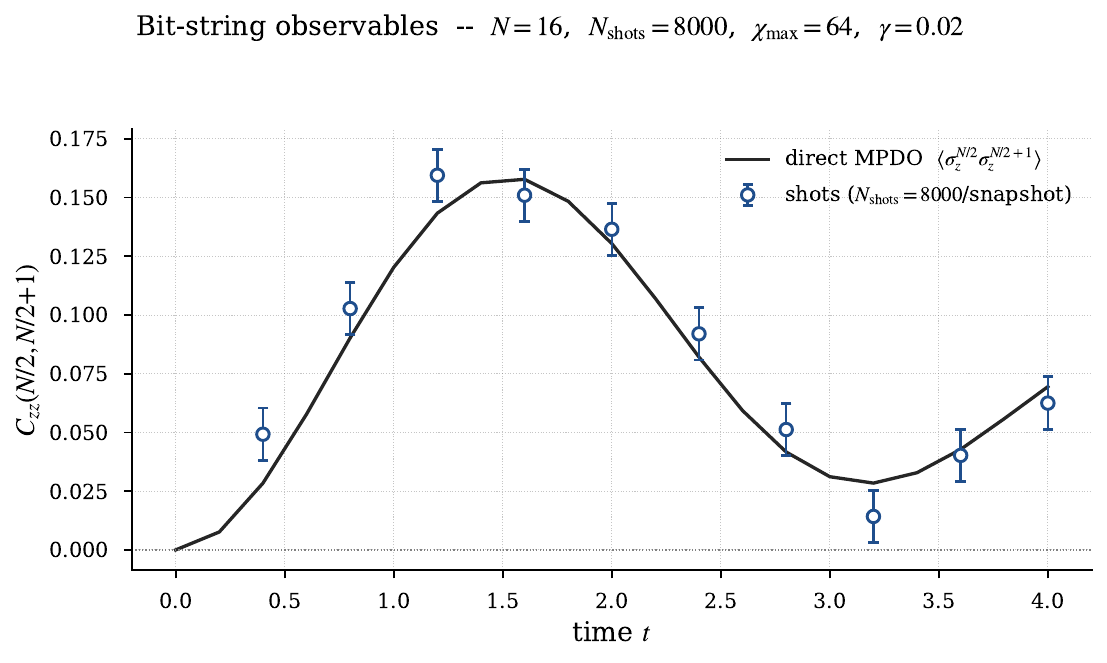}
\caption{Two-point $C_{zz}(N/2, N/2{+}1)$ reconstructed from perfect bitstring sampling of the MPDO (blue open circles, $N_{\rm shots}=8000$ shots per snapshot, $1\sigma$ s.e.m. error bars) compared against direct MPDO contraction (black line). Parameters: $N=16$, $\chi_{\max}=64$, $\gamma=0.02$, depolarizing channel, $\ket{+}^{\otimes N}$ initial state.}
\label{fig:sup_perfect_sampling_czz}
\end{figure}

\begin{figure}[h!]
\centering
\includegraphics[width=0.95\columnwidth]{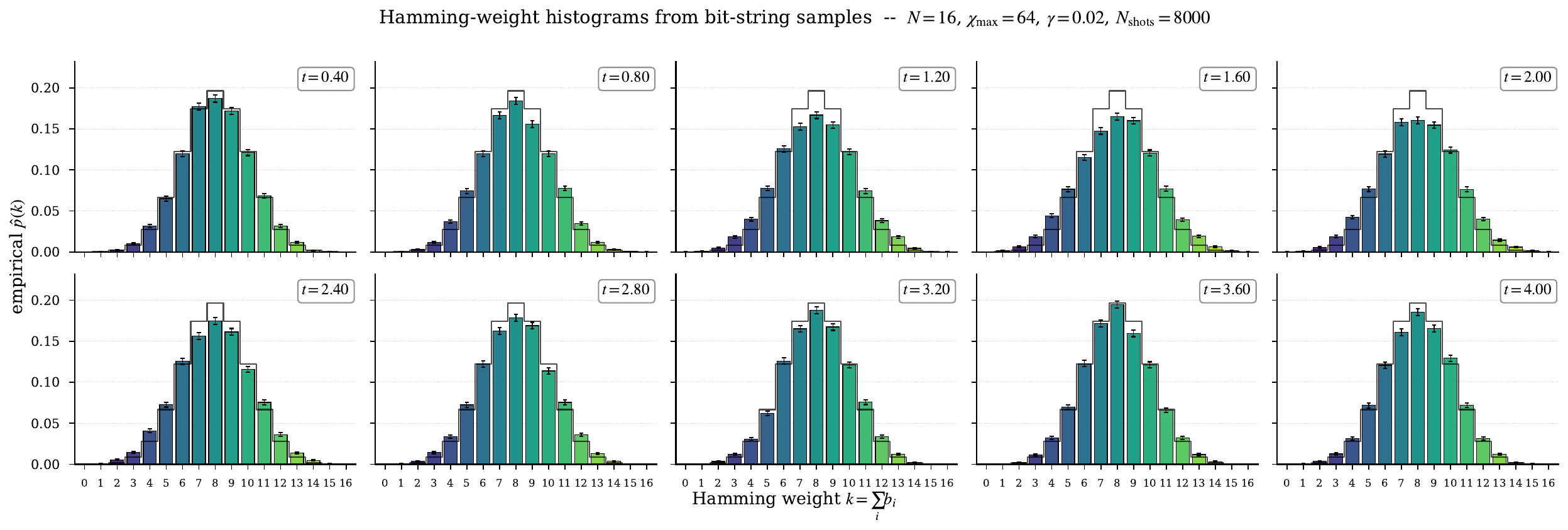}
\caption{Empirical Hamming-weight distributions $\hat p(k)$ from $N_{\rm shots}=8000$ perfect bitstring samples of the MPDO at ten snapshot times $t=0.4,0.8,\ldots,4.0$. Bars are coloured by Hamming weight $k$; error bars are binomial $1\sigma$. The thin black step line is the $\mathrm{Binom}(N,1/2)$ reference, i.e.\ the maximally-mixed Z-diagonal limit. Same parameters as Fig.~\ref{fig:sup_perfect_sampling_czz}.}
\label{fig:sup_perfect_sampling_hamming_grid}
\end{figure}

}
\end{document}